\documentclass[%
reprint,
superscriptaddress,
 amsmath,amssymb,
 aps,
prb,
]{revtex4-2}

\usepackage {graphicx,epsfig,graphics,color}
\usepackage{dcolumn}
\usepackage{bm}
\usepackage{hyperref}
\usepackage{sidecap}
\usepackage{float}

\begin{document}
\title{Collective magnetic Higgs excitation in a pyrochlore ruthenate}

\author{Dirk Wulferding}
\altaffiliation{Contributed equally to this work.}
\affiliation{Center for Correlated Electron Systems, Institute for Basic Science, Seoul 08826, Korea}
\affiliation{Department of Physics and Astronomy, Seoul National University, Seoul 08826, Korea}

\author{Junkyoung Kim}
\altaffiliation{Contributed equally to this work.}
\affiliation{Department of Physics, Incheon National University, Incheon 22012, Korea}

\author{Mi Kyung Kim}
\affiliation{Center for Correlated Electron Systems, Institute for Basic Science, Seoul 08826, Korea}
\affiliation{Department of Physics and Astronomy, Seoul National University, Seoul 08826, Korea}

\author{Yang Yang}
\affiliation{School of Physics and Astronomy, University of Minnesota, Minneapolis, Minnesota 55455, USA}

\author{Jae Hyuck Lee}
\affiliation{Center for Correlated Electron Systems, Institute for Basic Science, Seoul 08826, Korea}
\affiliation{Department of Physics and Astronomy, Seoul National University, Seoul 08826, Korea}

\author{Dongjoon Song}
\affiliation{Center for Correlated Electron Systems, Institute for Basic Science, Seoul 08826, Korea}
\affiliation{Department of Physics and Astronomy, Seoul National University, Seoul 08826, Korea}

\author{Dongjin Oh}
\affiliation{Center for Correlated Electron Systems, Institute for Basic Science, Seoul 08826, Korea}
\affiliation{Department of Physics and Astronomy, Seoul National University, Seoul 08826, Korea}

\author{Heung-Sik Kim}
\affiliation{Department of Physics, Kangwon National University, Chuncheon 24311, Korea}

\author{Li Ern Chern}
\affiliation{Department of Physics, University of Toronto, Toronto, ON M5S 1A7, Canada}

\author{Yong Baek Kim}
\affiliation{Department of Physics, University of Toronto, Toronto, ON M5S 1A7, Canada}

\author{Minji Noh}
\affiliation{Department of Physics and Astronomy, Seoul National University, Seoul 08826, Korea}
\affiliation{Institute of Applied Physics, Seoul National University, Seoul 08826, Korea}

\author{Hyunyong Choi}
\affiliation{Department of Physics and Astronomy, Seoul National University, Seoul 08826, Korea}
\affiliation{Institute of Applied Physics, Seoul National University, Seoul 08826, Korea}

\author{Sungkyun Choi}
\affiliation{Center for Integrated Nanostructure Physics, Institute for Basic Science, Suwon 16419, Korea}
\affiliation{Sungkyunkwan University, Suwon 16419, Korea}

\author{Natalia B. Perkins}
\affiliation{School of Physics and Astronomy, University of Minnesota, Minneapolis, Minnesota 55455, USA}

\author{Changyoung Kim}
\email[]{changyoung@snu.ac.kr}
\affiliation{Center for Correlated Electron Systems, Institute for Basic Science, Seoul 08826, Korea}
\affiliation{Department of Physics and Astronomy, Seoul National University, Seoul 08826, Korea}

\author{Seung Ryong Park}
\email[]{abepark@inu.ac.kr}
\affiliation{Department of Physics, Incheon National University, Incheon 22012, Korea}

\date{\today}

\begin{abstract}

The emergence of scalar Higgs-type amplitude modes in systems where symmetry is spontaneously broken has been a highly successful, paradigmatic description of phase transitions, with implications ranging from high-energy particle physics to low-energy condensed matter systems. Here, we uncover two successive high temperature phase transitions in the pyrochlore magnet Nd$_2$Ru$_2$O$_7$ at $T_{\mathrm{N}} = 147$ K and $T^* = 97$ K, that lead to giant phonon instabilities and culminate in the emergence of a highly coherent excitation. This coherent excitation, distinct from other phonons and from conventional magnetic modes, stabilizes at a low energy of 3 meV. We assign it to a collective Higgs-type amplitude mode, that involves bond energy modulations of the Ru$_4$ tetrahedra. Its striking two-fold symmetry, incompatible with the underlying crystal structure, highlights the possibility of multiple entangled broken symmetries.

\end{abstract}

\maketitle

\section{Introduction}

The Higgs mechanism, first invoked to describe spontaneous symmetry breaking in superconductors, has since proven successful in describing masses of the W and Z bosons in the standard model~\cite{higgs-64}, in describing collective excitations in superconducting condensates and superfluids~\cite{shimano-20}, and in describing quantum magnets close to criticality~\cite{jain-17, ruegg-08}. In these scenarios, Higgs modes of scalar nature are a crucial component of spontaneous symmetry breaking. Very recently, axial Higgs modes have been discussed in the context of condensed matter physics, where the spontaneous and simultaneous breaking of different symmetries gives rise to the interplay of multiple order parameters~\cite{wang-22}. As axial Higgs modes may be highly relevant for models describing physics beyond the standard model, e.g., dark matter~\cite{marsh-16, franzosi-16}, correlated electron systems with entangled degrees of freedom can serve as excellent playgrounds to explore exotic, non-scalar amplitude-Higgs modes and test proposed models~\cite{wang-22}.

Among correlated electron systems, 4$d$ transition metal oxides (TMOs) provide exceptional platforms for exploring complex emergent phenomena due to their intricate interplay of spin and orbital degrees of freedom~\cite{BookCao,Takayama2021}. Examples include spin fractionalization into Majorana fermions due to bond-dependent exchange frustration in Kitaev honeycomb magnets~\cite{Kitaev2006,Jackeli2009,takagi-19,Takayama2021,Trebst2022}, unconventional superconductivity~\cite{mackenzie-17}, spin-ice states with emergent magnetic monopoles and unique quantum electrodynamics~\cite{Bramwell2001,Gingras2014,lefrancois-17, pace-21}, and materials with topologically non-trivial electronic and magnon band structures~\cite{Li2016, wang-20, Hwang2020}, to name a few. Particularly, the family of 4$d$ transition metal pyrochlore oxides exhibits a diverse array of exotic phenomena, driven by the interplay between geometrical frustration, electronic interactions, spin-orbit coupling, and lattice instabilities~\cite{gao-20, gardner-10}.

The diverse physical properties of these 4$d$ TMO pyrochlores with the general chemical formula $A_2$$M_2$O$_7$ are determined by the choice of $A$-site ions and transition metal ions $M$. The Mott insulators $A_2$Ru$_2$O$_7$ with a series of rare-earth elements ($A$ = Pr -- Lu, Y), in which Ru$^{4+}$ ions (electron configuration 4$d^4$) are expected to carry $S=1$ spins and angular moment $L = 1$ according to Hund's rule, are of special interest as candidate materials to study magnetic properties of frustrated spin-1 antiferromagnetism~\cite{haldane-83, wang-15, li-18, buessen-18}. Particularly, a number of different spin structures both with and without topological magnons may be realized in this class of materials, depending on microscopic details~\cite{gao-20}. Yet, the lack of high quality single crystals of the $A_2$Ru$_2$O$_7$ series has hindered a systematic experimental exploration of their exact ground states and emergent magnetic anisotropies so far. Additional complications arise from the fact that spin-orbit coupling is comparable to the super-exchange scale. Therefore, a magnetic condensation of Van Vleck excitons may occur~\cite{Khaliullin2013, jain-17}, which will behave as the magnetic degrees of freedom and lead to either spin-nematic or amplitude fluctuations.

The compound of our interest, Nd$_2$Ru$_2$O$_7$, consists of Ru$^{4+}$ ions octahedrally coordinated by O$^{2-}$ ions to form larger Ru tetrahedra building blocks. Below $T_{\mathrm{N}} = 147$ K, Ru magnetic moments order antiferromagnetically (see Supplementary Note 1, and Supplementary Figures 1 and 2), although a dispute exists about the nature of the ordering -- while some experiments point towards a long-range ordered state~\cite{gaultois-13}, an earlier neutron diffraction study on powder samples of Nd$_2$Ru$_2$O$_7$ suggested a spin-glass or short-range ordered state~\cite{ito-00}.

Here, we employ polarization-resolved Raman spectroscopy to study the symmetry and the temperature-dependence of magnetic and phonon excitations in high quality Nd$_2$Ru$_2$O$_7$ single crystals. Below the N\'{e}el temperature $T_{\mathrm{N}} = 147$ K of the Ru ions we uncover an extended temperature regime of spin-lattice coupled fluctuations that result in a second phase transition $T^* = 97$ K and give rise to an anomalous, well-defined low-energy amplitude excitation. Our observations suggest that at intermediate temperatures Nd$_2$Ru$_2$O$_7$ is dominated by bond energy fluctuations from which a collective excitation emerges, which is consistent with a Higgs-type nematic amplitude mode.

\section{Results}

\subsection{Magnetic excitations}

\begin{figure*}
\label{figure1}
\centering
\includegraphics[width=16cm]{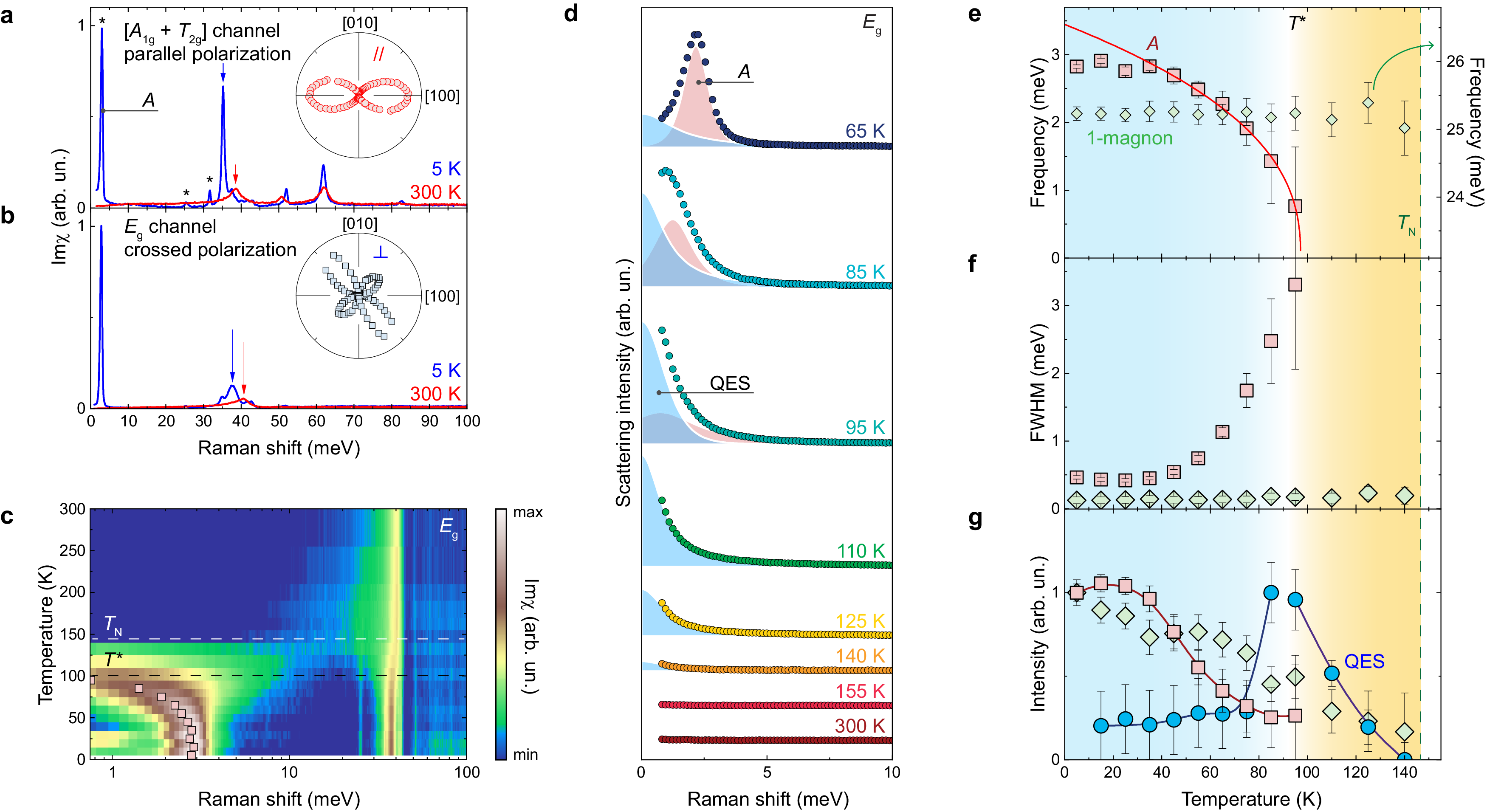}
\caption{\textbf{Temperature dependence of magnetic modes.} \textbf{a} and \textbf{b} Bose-corrected Raman intensity (Im$\chi$) measured at 5 K (blue curve) and 300 K (red curve) in the $A_{\mathrm{1g}}$+$T_{\mathrm{2g}}$ ($\textbf{e}_{\mathrm{in}}=\textbf{e}_{\mathrm{out}}=[110]$) and the $E_{\mathrm{g}}$ ($\textbf{e}_{\mathrm{in}}=[110],\textbf{e}_{\mathrm{out}}=[\bar{1}10]$) symmetry channels, respectively. Arrows mark the $T_{\mathrm{2g}}$ phonon (\textbf{a}) and the $E_{\mathrm{g}}$ phonon (\textbf{b}). The insets plot the angular dependent intensity of excitation $A$ at 5 K in parallel and crossed polarization. \textbf{c} Color contour plot of the temperature dependence in $E_{\mathrm{g}}$ symmetry. Note that the Raman shift is shown on a logarithmic scale to emphasize low lying excitations. Square symbols trace the evolution of excitation $A$. The dashed lines mark $T_{\mathrm{N}}$ and $T^*$. \textbf{d} Raman spectra measured in $E_{\mathrm{g}}$ symmetry at selected temperatures through $T_{\mathrm{N}}$ (full circles). The emergent quasi-elastic scattering (QES) and excitation $A$ are shaded. \textbf{e}-\textbf{g} Temperature dependence of the energy, linewidth, and intensity of the low-energy excitation $A$ (red squares) together with the one-magnon excitation at 25 meV (pale green diamonds). The solid red curve is a mean-field fit to the energy (see text for details). The blue circles in (\textbf{g}) show the intensity of quasi-elastic scattering. The dashed green line marks $T_{\mathrm{N}}$. Standard deviations in (\textbf{e-g}) are indicated by error bars.}
\end{figure*}

Below 1.8 K, Nd$^{3+}$ spins are found to order in an all-in-all-out state~\cite{ku-18}. In our study we only focus on temperatures above 5 K, therefore we can consider the Nd spins to be paramagnetic~\cite{gaultois-13} and thus assign all magnetic excitations in our study to Ru ions. In Figs. 1a and 1b we plot Bose-corrected Raman data obtained at 300 and 5 K in two different scattering geometries, distinguishing between the three allowed symmetry channels $A_{\mathrm{1g}}$+$T_{\mathrm{2g}}$ and $E_{\mathrm{g}}$ (see also Supplementary Note 2 and Supplementary Figure 3). All spectral features in the energy range 33 -- 90 meV can be assigned to phonon modes, which we will discuss in more detail below. Three low-energy excitations emerge below $T_{\mathrm{N}}$ at 32, 25, and 3 meV, marked by asterisks in Fig. 1a.

We can assign the two higher-lying excitations at 32 and 25 meV to one-magnon modes, based on their symmetry, their energies (see Supplementary Note 3 and Supplementary Figure 4), and their temperature dependence discussed below. Based on our analysis of symmetry-allowed spin configurations on the $S=1$ pyrochlore lattice, the best description of the 32 and 25 meV one-magnon modes in terms of energy and number of branches at \textbf{q}=0 is achieved for the all-in-all-out ordered state of Ru magnetic moments. This result is somewhat at odds with previous powder neutron diffraction results~\cite{ito-00}, where a spin-glass state or XY spin structure was assigned to Nd$_2$Ru$_2$O$_7$. Our assignment is based on experiments restricted to \textbf{q}=0. Furthermore, additional terms added to the Hamiltonian, such as substantial spin-orbit coupling, may modify this scenario and other spin configurations may be better candidates to describe our experimental data. Eventually, a detailed temperature-dependent inelastic neutron scattering study on high-quality phase-pure single crystalline Nd$_2$Ru$_2$O$_7$ will help to settle this issue.

Phenomenologically similar one-magnon excitations have been reported in the spin-1/2 pyrochlore compound Y$_2$Ir$_2$O$_7$~\cite{nguyen-21}, as well as in recent Raman scattering experiments from powder samples of $A_2$Ru$_2$O$_7$ ($A$ = Y, Sm, Eu, Ho, Er)~\cite{jaehyeok-22}. In addition, the overall small intensities of these modes relative to other phonon modes can support our assignment, as magnetic Raman scattering processes are generally less intense than phononic Raman scattering processes.

In contrast to these one-magnon modes of weak scattering intensities, the peak at 3 meV (in the following referred to as the $A$-mode) dominates the Raman spectra at low temperatures. The angular dependence of its intensity, measured at $T = 5$ K, is plotted as a function of light polarization within the $ab$ plane with parallel polarization (inset of Fig. 1a) and crossed polarization (inset of Fig. 1b). Its two-fold symmetry in the parallel polarization and the modulation of the lobe's amplitudes in the cross polarization are incompatible with expected $A_{1}$, $T_{2}$, or $E$ irreducible representations of the $T_{\mathrm{d}}$ lattice point group of the pyrochlore lattice, and require a Raman tensor with antisymmetric tensor elements. Therefore these results are also different from the four-fold symmetries expected and observed for conventional (i.e., one-magnon or two-magnon) magnetic modes (see Supplementary Notes 4, 5, and Supplementary Figure 5 for a detailed symmetry analysis).

To better understand the nature of this unconventional excitation, we now turn to an analysis of its thermal evolution. The temperature dependent Raman data is shown in a color contour plot in Fig. 1c (see Supplementary Note 6 and Supplementary Figure 6 for details on the fitting procedure). As we approach $T_{\mathrm{N}}$ from the high temperature side, there is a continuous increase in a broadened spectral weight at the left shoulder of the $E_{\mathrm{g}}$ phonon around 40 meV (see the green-colored background between 10-40 meV). In agreement with previous Raman studies on pyrochlore magnets~\cite{ueda-19}, we assign this continuum to paramagnetic fluctuations. Once $T_{\mathrm{N}}$ is reached and crossed, this spectral weight is transferred towards zero energies, forming an increasing quasi-elastic scattering signal. Note that the spectral weight is not necessarily a conserved quantity through the transition process. Nonetheless, the transfer of spectral weight suggests an intimate link between the two signals. The quasi-elastic signal can be well fitted by a single Lorentzian line-shape centered at $E=0$, as demonstrated in Fig. 1d for the temperatures 110-140 K. Such behavior is characteristic for a phase transition and originates from increasing zero-energy density fluctuations of the order parameter, which diverge through a phase transition~\cite{gallais-16, lemmens-03, wang-20b, reiter-76, lyons-82, kim-21}.

Only below a second transition temperature $T^* = 97$ K does the quasi-elastic scattering intensity decrease, giving rise to a massive excitation, marked by the pale red shaded area in Fig. 1d for a few selected temperatures. The peak position determined from the fit is shown with pale red squares in Fig. 1c. Interestingly, neither bulk magnetization, nor specific heat shows any hint of this phase transition $T^*$ (see Supplementary Figure 1 and Supplementary Note 1). Despite these missing fingerprints, the dispersion of the $A$ peak suggests that this mode may be linked to some bond spin-nematic order parameter $\eta$, which lowers the crystalline symmetry of the pyrochlore lattice below $T^*$.

The full temperature evolution of the $A$-mode is given in the panels of Figs. 1e-g, where we analyze frequency, line width, and intensity of this excitation. Fig. 1g additionally plots the intensity over temperature of the quasi-elastic signal for direct comparison. Comparing the thermal evolution of the $A$-mode (red squares) with that of the 25 meV one-magnon mode (green diamonds), we notice a fundamentally different behavior. The nearly temperature-independent energy of the one-magnon excitation stands in stark contrast to the dramatic softening of mode $A$, which instead follows an order-parameter-like behavior that can be described by a critical exponent, i.e., $\omega(T) = \alpha \,\vert T-T^* \vert^{\beta}$, with $\alpha$ being a scaling factor, $T^* = (97 \pm 5)$ K the critical temperature, and $\beta = 0.4 \pm 0.1$ the critical exponent. Likewise, while the linewidth of the one-magnon mode at 25 meV remains close to constant over a wide temperature range and up to $T_{\mathrm{N}}$, the width of the $A$-mode broadens significantly with increasing temperature. Finally, the intensity of the 25-meV mode gradually drops off and approaches 0 around $T_{\mathrm{N}}$, indicating a decreasing magnetic moment of the Ru ions upon approaching $T_{\mathrm{N}}$. The $A$-mode, instead, evidences a quick drop in intensity with increasing temperature and disappears at $T^*$. As detailed in Supplementary Note 7 and Supplementary Figure 7, the thermal evolution of the 32 meV excitation mirrors that of the 25-meV one.

Considering all of the above, we can confidently rule out a one-magnon scattering process to be related to the $A$-mode. Likewise, a purely phononic origin of this mode is ruled out, since preliminary neutron scattering data on powder samples shows no hint of a phonon mode at 3 meV at base temperature [S. Choi, private communication]. We also emphasize that neither optical phonons, nor zone-folded acoustic phonons would be expected at such low energies in pyrochlore systems with this or similar elemental composition~\cite{kim-20}. Zone-folding effects can be further discarded by taking the dominating spectral weight of the $A$-mode into account, which would require a significant degree of lattice anharmonicity. No other structural refinement method however has lead to the observation of such dramatic structural effects. In addition, the rather sharp linewidth of the mode at base temperature underlines the coherent nature of the corresponding excitations, which agrees with the picture of fluctuation of some magnetic order parameter.

\subsection{Lattice dynamics and its connection to the bond nematic amplitude mode}

\begin{figure*}
\label{figure2}
\centering
\includegraphics[width=10cm]{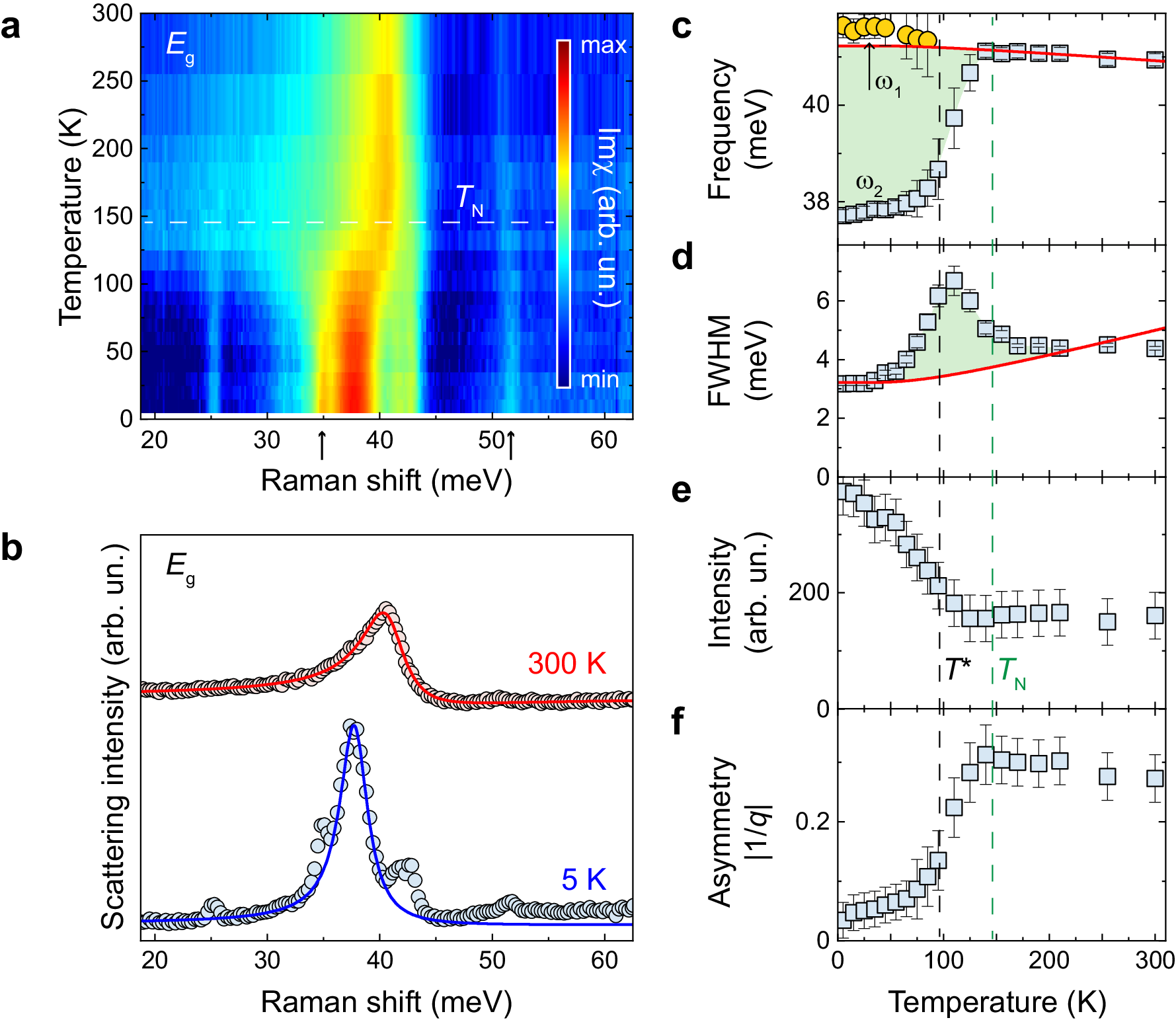}
\caption{\textbf{Giant spin-phonon coupling in Nd$_2$Ru$_2$O$_7$.} \textbf{a} Color contour plot of the mid-energy range focusing on the thermal evolution of the $E_{\mathrm{g}}$ channel. Arrows mark leakage of phonons from the $T_{\mathrm{2g}}$ channel. \textbf{b} Phonon spectra of the $E_{\mathrm{g}}$ channel measured at 5 K and at 300 K (symbols) together with asymmetric Fano fits to the dominating $E_{\mathrm{g}}$ phonon mode (solid lines). \textbf{c}-\textbf{f} Temperature-dependence of the phonon parameters frequency, linewidth (FWHM), peak intensity, and Fano asymmetry for the $E_{\mathrm{g}}$ phonon. The solid red lines are fits to the frequency and linewidth, corresponding to anharmonic softening and broadening, respectively (see text for details). Deviations from this anharmonic behavior are shaded in pale-green. $T_{\mathrm{N}}$ and $T^*$ are marked by dashed lines. Standard deviations in (\textbf{c-f}) are indicated by error bars.}
\end{figure*}

Evidence for the $A$-mode arising through fluctuations of a nematic order parameter can be also found in the lattice dynamics. To uncover such spin-related phonon anomalies, we now investigate the enhanced lattice dynamics in Nd$_2$Ru$_2$O$_7$ which we observe down to about half of $T_{\mathrm{N}}$. As the $E_{\mathrm{g}}$ phonon involves twisting motions of the RuO$_6$ octahedra, it is rather susceptible to the changes in the Ru-O-Ru bond angle. Therefore, it justifies a strong spin-elastic coupling between the $E_{\mathrm{g}}$ phonon and Ru magnetic moments, and such coupling serves as a sensitive local probe to the onset of magnetic order or even to an enhancement of spin-spin correlations within Ru tetrahedra. In Fig. 2a we present the mid-energy range of the $E_{\mathrm{g}}$ symmetry channel, which is dominated by the $E_{\mathrm{g}}$ phonon at 40 meV. The arrows at 35 and 52 meV mark small leakage of phonons from the $T_{\mathrm{2g}}$ channel. Below $T_{\mathrm{N}}$ a giant softening of the phonon frequency is observed, followed by a highly dynamical, fluctuating regime in which the phonon splits into a dominating low-energy mode ($\sim 38$ meV) and a weaker high-energy shoulder ($\sim 41$ meV). Below $T^*$, the lattice dynamics resumes its rather static behavior.

Fig. 2b highlights the lineshape of the $E_{\mathrm{g}}$ phonon, which is highly asymmetric and can be described by a Fano lineshape at room temperature. The 5 K spectrum evidences a nearly symmetric Lorentzian phonon lineshape. The appearance of an asymmetric Fano lineshape is generally ascribed to the interference of a discrete (phonon) mode with an underlying, broad continuum of excitations~\cite{fano-61}. A similar temperature dependence has been observed in the all-in-all-out pyrochlore compound Cd$_2$Os$_2$O$_7$~\cite{nguyen-17}, where it was related to a metal-insulator transition driven by spin-charge-lattice coupling. We recall that Nd$_2$Ru$_2$O$_7$ remains electrically insulating across the investigated temperature regime~\cite{gaultois-13}, therefore we rule out an electronic continuum as the origin (for details on the electronic band structure, see~\cite{elstructure}). Instead, incoherent spin fluctuations of the paramagnetic phase ($T>T_{\mathrm{N}}$) are a natural candidate for the broad continuum, as reported in the pyrochlore iridate Eu$_2$Ir$_2$O$_7$~\cite{ueda-19}. Upon passing through $T_{\mathrm{N}}$, the spectral weight of the continuum is transferred towards quasi-elastic scattering and eventually to the low-energy $A$-mode. Thus, below $T^*$ the interference between the broad magnetic continuum and the $E_{\mathrm{g}}$ phonon mode disappears and the phonon approaches a symmetric Lorentzian lineshape at low temperatures.

An analysis of the phonon peak parameters frequency, linewidth, intensity, and Fano asymmetry is presented in Figs. 2c-f. Both frequency and linewidth can be described by anharmonic behavior for $T>T_{\mathrm{N}}$ (see solid red line and Supplementary Note 8 for details)~\cite{balkanski-83}. Below $T_{\mathrm{N}}$, this anharmonic behavior is abruptly halted and replaced by an enormous energy softening, peak broadening, and line splitting. A softening of the $E_{\mathrm{g}}$ phonon with the onset of magnetic order is commonly observed in pyrochlore magnets~\cite{lee-04, thomas-22, son-19}. The giant softening from 41 meV down to 38 meV, present in Nd$_2$Ru$_2$O$_7$, however, exceeds these common values by about one order of magnitude. The dramatically enhanced linewidth in between the transition temperatures signals the opening of an effective decay channel in the form of additional spin fluctuations. Simultaneously, the intensity of the dominating line starts to increase monotonically, and the Fano asymmetry drops and approaches zero towards lowest temperatures. All these phonon anomalies underline the existence of a highly dynamical lattice that strongly couples to spin degrees of freedom, which settle in modulated or canted spin structures and culminate in the formation of an amplitude mode, i.e., the $A$-mode at 3 meV.

\subsection{Broken symmetries}

\begin{figure*}
\label{figure3}
\centering
\includegraphics[width=10cm]{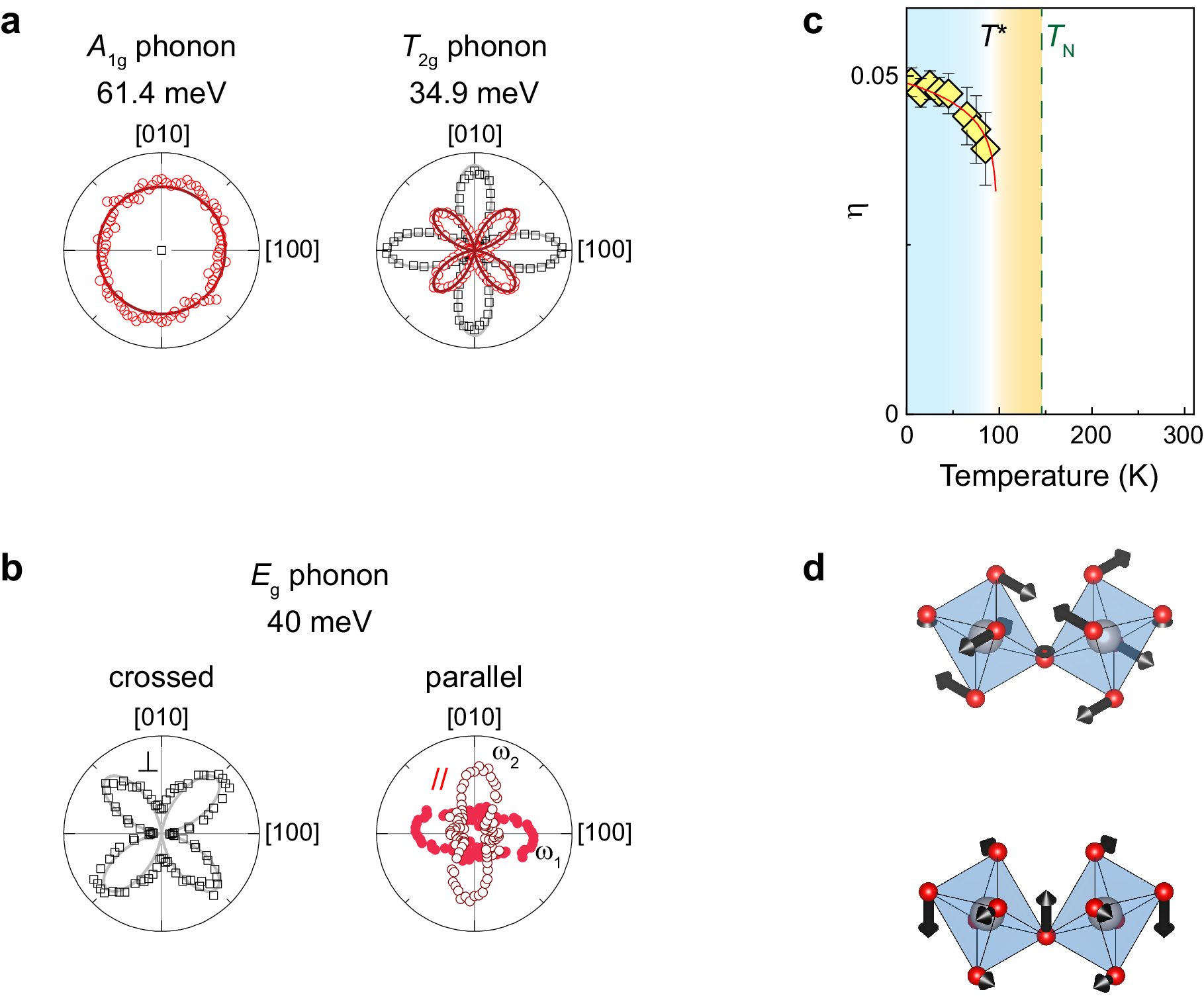}
\caption{\textbf{Evidence for symmetry breaking in Nd$_2$Ru$_2$O$_7$.} \textbf{a} Polarization plots for phonons of $A_{\mathrm{1g}}$, $T_{\mathrm{2g}}$, and \textbf{b} $E_{\mathrm{g}}$ symmetry measured in parallel (red circles) and in crossed (black squares) configuration at $T = 5$ K. The solid red and gray lines trace the theoretical rotational anisotropy following the respective Raman tensors. \textbf{c} The nematic order parameter $\eta (T)$ as extracted from the phonon frequencies shown in Fig. 2c. \textbf{d} Arrows sketch the ionic displacement patterns of the doubly degenerate $E_{\mathrm{g}}$ mode. Standard deviations in (\textbf{c}) are indicated by error bars.}
\end{figure*}

With the dynamic aspect of various Raman modes, we now turn to an analysis of the symmetry properties of various excitations in Nd$_2$Ru$_2$O$_7$. As we already detailed in Figs. 1a and 1b, the $A$-mode displays a striking two-fold symmetry in both parallel and crossed scattering configurations, which is at odds with the given cubic crystal structure. If the anomalous lattice dynamics of the $E_{\mathrm{g}}$ mode is indeed from coupling to magnetic bond-energy fluctuations, there should be some evidence of this coupling in its symmetry properties as well. Therefore, we now investigate the symmetries of the phonon modes.

Fig. 3a shows polar plots of phonon intensities of different symmetries, measured at $T = 5$ K in parallel ($xx$, red circles) and crossed ($xy$, black squares) polarization (see Supplementary Figure 3 for the full data set). The experimental data for the $A_{\mathrm{1g}}$ and $T_{\mathrm{2g}}$ modes match their fits very well, which are denoted by solid lines based on the respective Raman tensors given in Supplementary Note 2. In contrast, the splitting of the $E_{\mathrm{g}}$ phonon mode into $\omega_1$ and $\omega_2$ lowers the symmetry of the resulting branches observed in parallel polarization from four-fold to two-fold rotationally symmetric (Fig. 3b), thus pointing towards a lowering symmetry of the RuO$_6$ octahedra. Without structural distortion, such lowering symmetry can be triggered from a spin bond order parameter $\langle\mathbf{f}_\eta\rangle=(\langle f_1\rangle,\langle f_2\rangle)$, where $f_1=[(\mathbf{S}_1+\mathbf{S}_2)\cdot(\mathbf{S}_3+\mathbf{S}_4)-2\mathbf{S}_1\cdot\mathbf{S}_2-2\mathbf{S}_3\cdot\mathbf{S}_4]/\sqrt{12}$ and $f_2=(\mathbf{S}_1-\mathbf{S}_2)\cdot(\mathbf{S}_3-\mathbf{S}_4)/2$, as proposed in~\cite{Tchernyshyov2002}. Such spin bond order parameter measures bond energy differences within the Ru tetrahedron, and couples directly to the $E_{\mathrm{g}}$ phonon mode through the spin-elastic coupling $\mathbf{f}_\eta\cdot \mathbf{u}_{E_{\mathrm{g}}}$. In crossed polarization, both branches still appear with the same four-fold symmetry (see Fig. 3b and Supplementary Figure 3]. Both the splitting and the symmetry reduction are reminiscent of a nematic-like phase transition~\cite{yao-22}. We can rule out any significant misalignment of the crystal or contributions from (potentially existing) neighboring domains as sources for these distorted polar plots, since all other phonons closely follow their expected rotational symmetry and only the $E_{\mathrm{g}}$ mode is affected.

In Fig. 3c we quantify the energy difference between $\omega_1$ and $\omega_2$ by introducing $\eta = (\omega_1 - \omega_2)/(\omega_1 + \omega_2)$, a phonon anisotropy parameter~\cite{zhang-16}, which can be associated with the spin bond order parameter $\langle\mathbf{f}_\eta\rangle$, and we treat $\eta$ as the nematic order parameter. Fig. 3d sketches the ionic displacement patterns corresponding to the two-fold degenerate $E_{\mathrm{g}}$ phonon mode. Upon distortion of the local octahedral environment within a unit cell, this two-fold degeneracy may be lifted, resulting in a splitting of the $E_{\mathrm{g}}$ phonon. Previous studies on Nd$_2$Ru$_2$O$_7$ have not found any indication of a structural phase transition within our temperature range of interest, with experimental methods ranging from synchrotron and x-ray diffraction~\cite{gaultois-13, chen-15}, to neutron diffraction~\cite{ku-18}. However, a pronounced, non-monotonic decrease in the Ru-O-Ru angle was observed between 150 K -- 100 K~\cite{chen-15}. This can indicate a gradual canting or rearrangement of spins in a (short-ranged) antiferromagnet or a transition among two neighboring phases with decreasing temperature, accompanied by strong bond-energy fluctuations and manifested by the observed entangled spin-lattice instabilities. This concept of competing bond configurations has also been explored, e.g., in breathing pyrochlores~\cite{lee-21, dissanayake-22}, while spin-lattice coupling in a 2D magnet lead to the splitting of an $E_{\mathrm{g}}$ phonon, albeit of smaller magnitude~\cite{tian-16}. The temperature dependence of $\eta$ closely follows that of the $A$-mode, hence its emergence below 97 K indicates the onset of nematic order. Also, the fact that the two-fold, distorted $E_{\mathrm{g}}$ polar plot resembles the rotational symmetry of excitation $A$, as shown in Figs. 1a and 1b, suggests a direct correlation between the amplitude $A$-mode and lattice degrees of freedom of the RuO$_6$ octahedra.

\section{Discussion}

With the broken symmetry identified, we wish to discuss the origin of the $A$-mode. In the related antiferromagnet Ca$_2$RuO$_4$~\cite{jain-17, souliou-17} a broader Higgs mode at high energies (centered around 40 meV) is reported to stabilize through a condensation of $J=1$ excitons per Ru site~\cite{Khaliullin2013}. This single-ion picture may however not be relevant for the stabilization of the $A$-mode in Nd$_2$Ru$_2$O$_7$, given its smaller energy scale and the fact that it emerges through a second phase transition within the magnetically-ordered phase. Instead, narrow spectral features at low energies with dominating scattering intensity -- strikingly similar to the $A$-mode -- have been observed via Raman spectroscopy and interpreted as Higgs-type amplitude modes in the charge-density-wave compounds GdTe$_3$~\cite{wang-22}, 2$H$-TaS$_2$~\cite{grasset-19}, and NbSe$_2$~\cite{measson-14}.

Based on these considerations and on the observed lattice dynamics, we can pinpoint the following scenario for the formation of the $A$-mode: At $T_{\mathrm{N}}$, when the order of the Ru magnetic moments develops and translational symmetry breaks, the spin bond order parameter $\langle \mathbf{f}_\eta\rangle$ remains zero, since the magnetic energy of the six bonds in a single tetrahedron is uniform. However, through the spin-elastic coupling to the $E_{\mathrm{g}}$ phonon, the magnetic energy of the system can be further lowered. This drives the development of the spin-nematic bond order, an example of vestigial order which grows out of the magnetic fluctuations, and in the temperature range between 97 K and 147 K leads to strong quasi-elastic scattering (Fig. 1d). Below 97 K, the energy of the $A$-mode grows as an order parameter (Fig. 1c). This order breaks the discrete rotational symmetry since it leads to inequality between bond energies on the Ru tetrahedron but preserves the translational symmetry, i.e., it is a \textbf{q}=0 order. The consequences of the spin nematic order also have a feedback to the lattice, causing the splitting of the $E_{\mathrm{g}}$ phonon with reduced two-fold symmetry through the spin-elastic coupling $\mathbf{f}_\eta\cdot \mathbf{u}_{E_{\mathrm{g}}}$. Within this picture, $T^*$ is the spin-nematic transition further lowering the symmetry of the system.

Remarkably, a two-fold symmetric amplitude Higgs mode was recently reported in the charge-density-wave ordered GdTe$_3$ in Raman scattering experiments of both parallel and crossed polarization configurations~\cite{wang-22}. As such low symmetry is also inconsistent with the scalar nature of the Higgs mode, a polarization-dependent mechanism of constructive / destructive pathway interference was invoked, which instead yields an exotic axial Higgs mode. The striking similarities between GdTe$_3$ and Nd$_2$Ru$_2$O$_7$ raise the exciting prospect of stabilizing an axial Higgs mode of magnetic nature in pyrochlore systems with multiple spontaneously broken symmetries, e.g., both magnetic and nematic order. A deeper insight into the spin dynamics and requires a temperature-dependent study of the magnon band structure, e.g., via inelastic neutron scattering or resonant elastic x-ray scattering experiments. In addition, second-harmonic generation studies may shine further light on the nature of broken symmetries. However, these methods require sizable single crystals with flat or cleavable surfaces. Given the availability of suitable single crystals, such future studies will be essential to unambiguously uncover the nature of the observed amplitude mode in Nd$_2$Ru$_2$O$_7$.

In summary, we uncover a wide temperature regime of coupled spin and lattice fluctuations in the pyrochlore ruthenate Nd$_2$Ru$_2$O$_7$ via Raman spectroscopy that result in a second phase transition $T^*$, which has been elusive in other thermodynamic probes. A low-energy excitation emerging out of these fluctuations with a distinct symmetry and a peculiar temperature evolution is interpreted as a coherent amplitude mode out of nematic order. Future studies under extreme conditions, such as high pressure and magnetic fields, are envisaged to tune these fluctuations and further pinpoint the nature of the low-energy Higgs mode.

\section{Methods}

\textbf{Sample Growth.} Single crystals of Nd$_2$Ru$_2$O$_7$ were synthesized using the KF flux method~\cite{millican-07}. Typical resulting crystals of about 30 $\times$ 30 $\times$ 30 $\mu$m$^3$ volume exhibit shiny triangular- and rectangular-shaped as-grown surfaces, corresponding to [111] and [100] facets, respectively. Based on our obtained Raman spectra and basic thermodynamic characterization [see Supplementary Note 1] we can rule out any inclusion of a secondary Nd$_3$RuO$_7$ phase, commonly found in pyrochlore ruthenates~\cite{gaultois-13, taira-99}.

\textbf{Raman Scattering.} Raman spectroscopic experiments have been carried out using a $\lambda = 532$ nm solid state laser (Cobolt Samba) with a spot diameter of about 10 $\mu$m and an incident laser power at the sample position below 0.35 mW to reduce local laser heating effects. The scattered light passed through a volume Bragg grating notch filter set (Optigrate) to discriminate the laser line and to access Raman signals with energies as low as 0.8 meV. The sample was mounted via silver epoxy onto the cold finger of a He-cooled open-flow cryostat (Oxford Microstat HiRes). The Raman-scattered light was dispersed through a single-stage Horiba iHR 320 spectrometer with a 1800 gr/mm grating onto a Horiba Synapse CCD. In this configuration, a spectral resolution of about 1.2 cm$^{-1}$ / pixel is achieved. A constant background (dark current) of 117 counts was subtracted from each spectrum. Where indicated, Bose-corrected Raman intensity Im$\chi(\omega)$ is plotted. It is related to the as-measured Raman intensity $I(\omega)$ via the fluctuation-dissipation theorem by $I(\omega) = [1+n(\omega)]$Im$\chi$, where $n(\omega)$ is the Bose factor.

\section{acknowledgments}
We acknowledge important discussions with Tae Won Noh, SungBin Lee and Giniyat  Khaliullin. This work was supported by the Institute for Basic Science (IBS) (Grant Nos. IBS-R009-G2, IBS-R009-Y3) and by the NRF (Grant No. 2020R1A2C1011439). M.N. and H.C. were supported by the National Research Foundation of Korea (NRF) through the government of Korea (Grant No. 2021R1A2C3005905), Scalable Quantum Computer Technology Platform Center (Grant No. 2019R1A5A1027055), Creative Materials Discovery Program (Grant No. 2017M3D1A1040828), and the Institute for Basic Science (Grant No. IBS-R034-D1). L.E.C. and Y.B.K. are supported by the NSERC of Canada. S.C. acknowledges support by the Institute for Basic Science (IBS-R011-Y3-2021). Y.Y. and N.B.P. were supported by the National Science Foundation under Award No. DMR-1929311.

\newpage

\textbf{Supplementary Note 1 $|$ Magnetic Susceptibility, Specific Heat, and Sample Quality}

In Supplementary Figure 1a we show the magnetization curves measured at a magnetic field of 100 Oe in zero-field-cooled and field-cooled mode. Due to the small size of individual crystals these measurements were performed on a polycrystalline pellet of Nd$_2$Ru$_2$O$_7$, thereby averaging over random field directions. Measurements of the specific heat of a polycrystalline pellet with and without applied magnetic field are shown in Supplementary Figure 1b. Both susceptibility and specific heat measurements mark a sharp, well-defined phase transition at the N\'{e}el temperature at 147 K, while lacking any signature for additional transitions around 100 K. We also notice the lack of an additional transition around 25 K. This is in contrast to several previous studies on Nd$_2$Ru$_2$O$_7$ powder samples, and points to the absence of a secondary Nd$_3$RuO$_7$ phase~\cite{gaultois-13, taira-99}.

\begin{figure*}
\label{figure4}
\centering
\includegraphics[width=12cm]{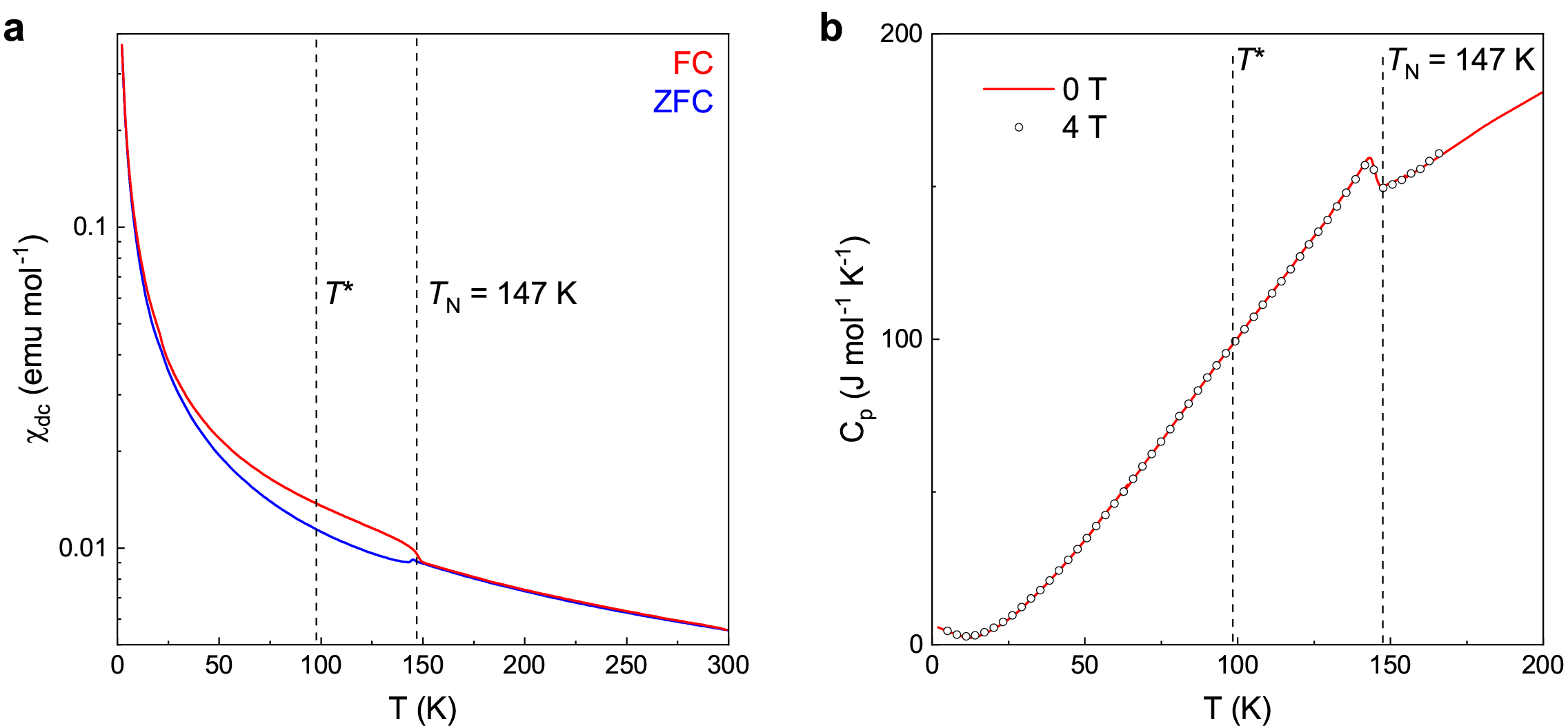}
\caption{\textbf{Thermodynamic characterization.} \textbf{a} Zero-field-cooled (blue) and field-cooled (red) magnetization measurements as a function of temperature at an applied magnetic field of 100 Oe. \textbf{b} Specific heat measurements at $B = 0$ T and $B = 4$ T.}
\end{figure*}

Our Raman scattering experiments performed on micrometer-sized Nd$_2$Ru$_2$O$_7$ single crystals with naturally-formed, shiny triangular (111-surface) and rectangular (100-surface) facets can help to further rule out any influence from a secondary 317 phase: The phonon spectra obtained on these samples are entirely consistent with the 227-pyrochlore structure and clearly incompatible with the secondary 317 phase [see Supplementary Figure 2].

\begin{figure*}
\label{figure5}
\centering
\includegraphics[width=5cm]{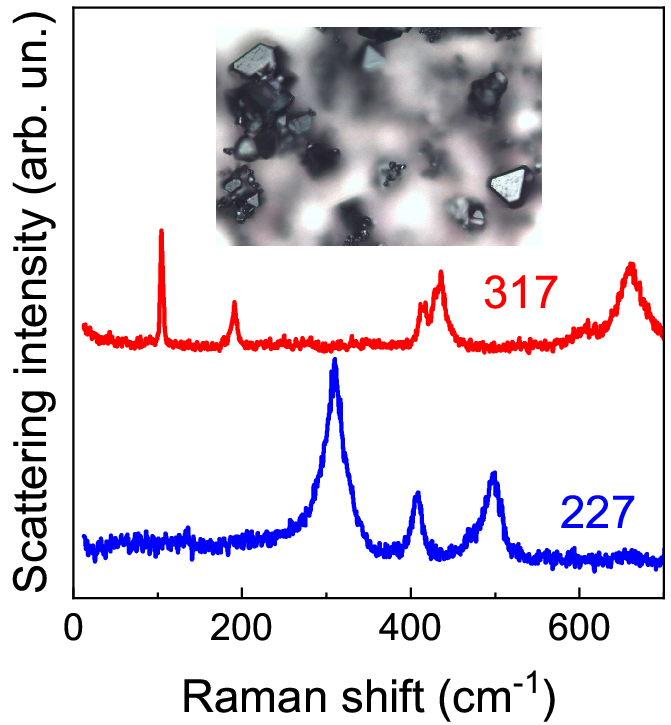}
\caption{\textbf{Sample Quality.} Room temperature Raman spectra of the 227 and 317 phases. Inset: Microscope image of crystallite clusters, including shiny as-grown surfaces.}
\end{figure*}

\textbf{Supplementary Note 2 $|$ Phonon Assignment and Selection Rules}

Nd$_2$Ru$_2$O$_7$ crystallizes in the cubic, centro-symmetric space group $Fd\bar{3}m$, with atoms occupying the following atomic positions: Nd -- (0.5, 0.5, 0.5), Ru -- (0, 0, 0), O1 -- (0.3309, 0.125, 0.125), O2 -- (0.375, 0.375, 0.375)~\cite{ku-18}. These correspond to the Wyckoff positions: Nd -- $16d$, Ru -- $16c$, O1 -- $8b$, O2 -- $48f$~\cite{ito-00}, which result in the six Raman-active modes $A_{\mathrm{1g}}$ + $E_{\mathrm{g}}$ + 4$T_{\mathrm{2g}}$, and their corresponding Raman tensors

\begin{center}
\mbox{$A_{\mathrm{1g}}$=$\begin{pmatrix} a & 0 & 0\\ 0 & a & 0\\ 0 & 0 & a\\
\end{pmatrix}$
, $E_{\mathrm{g}}^1$=$\begin{pmatrix} c & 0 & 0\\ 0 & c & 0\\ 0 & 0 & -2c\\ \end{pmatrix}$
, $E_{\mathrm{g}}^2$=$\begin{pmatrix} -\sqrt{3}c & 0 & 0\\ 0 & \sqrt{3}c & 0\\ 0 & 0 & 0\\ \end{pmatrix}$,}
\end{center}

\begin{center}
\mbox{$T_{\mathrm{2g}}^1$=$\begin{pmatrix} 0 & 0 & 0\\ 0 & 0 & d\\ 0 & d & 0\\ \end{pmatrix}$
, $T_{\mathrm{2g}}^2$=$\begin{pmatrix} 0 & 0 & d\\ 0 & 0 & 0\\ d & 0 & 0\\ \end{pmatrix}$
, $T_{\mathrm{2g}}^3$=$\begin{pmatrix} 0 & d & 0\\ d & 0 & 0\\ 0 & 0 & 0\\ \end{pmatrix}$}
\end{center}

\begin{figure*}
\label{figure6}
\centering
\includegraphics[width=12cm]{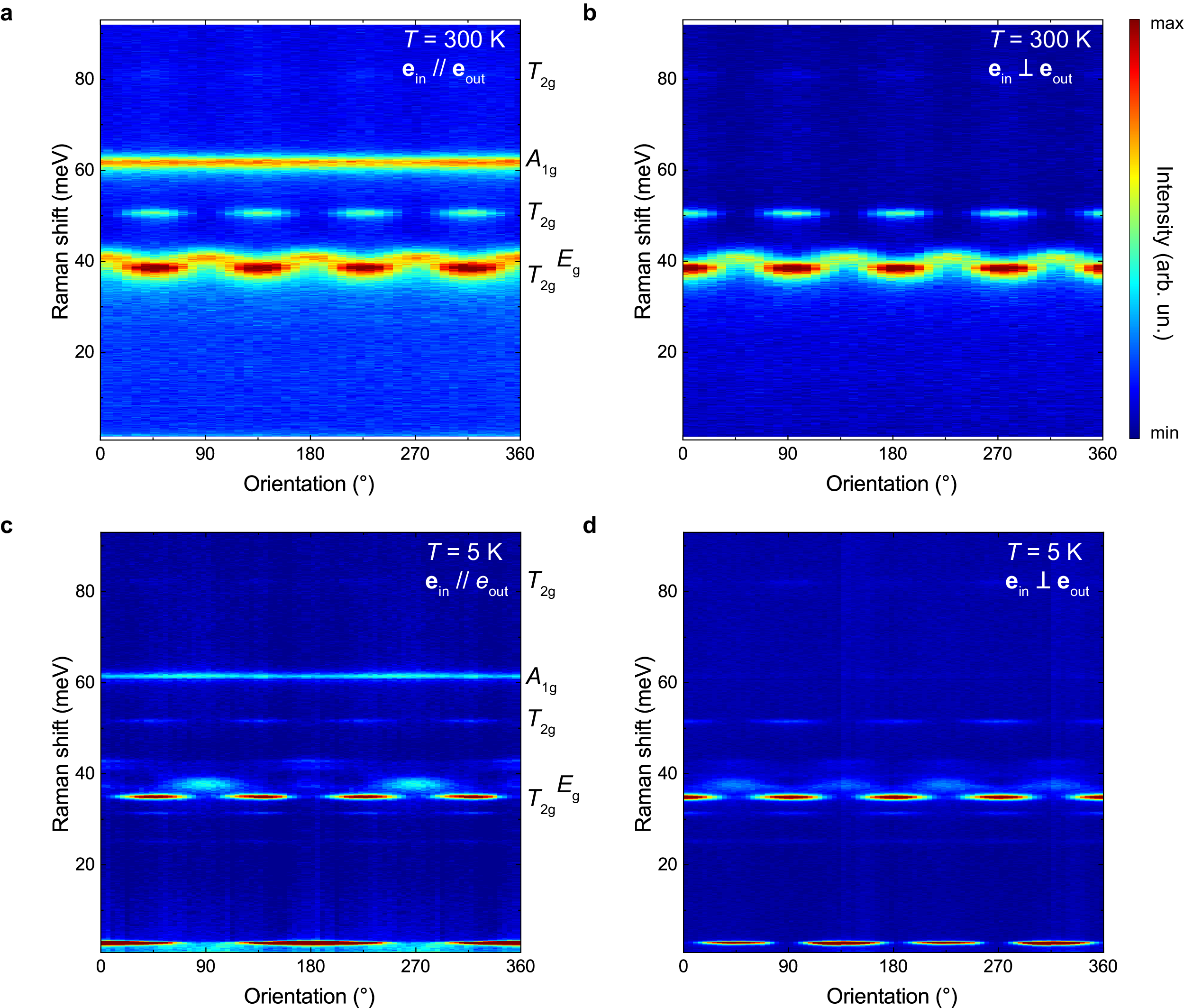}
\caption{\textbf{Polarization-resolved color contour plots.} Raman spectra taken on the [100] surface of Nd$_2$Ru$_2$O$_7$ at $T = 300$ K \textbf{a} in parallel and \textbf{b} in crossed polarization. \textbf{c} and \textbf{d} Corresponding color contour plots measured at $T = 5$ K.}
\end{figure*}

\begin{figure*}
	\label{figureMagnon}
	\centering
	\includegraphics[width=7cm]{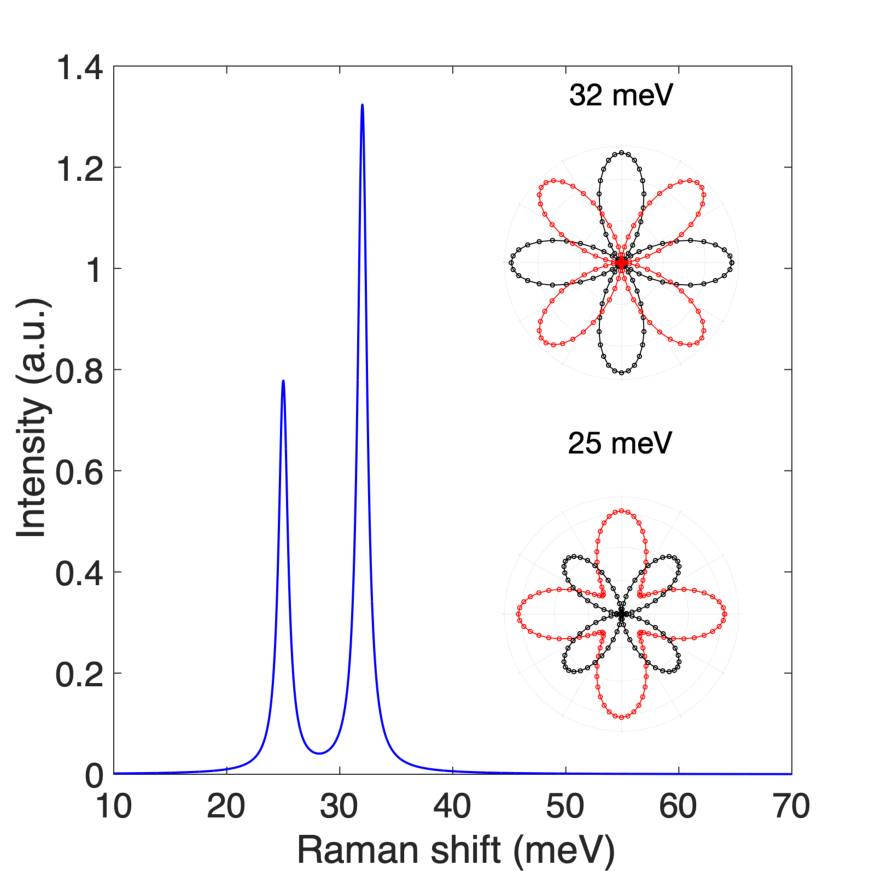}
	\caption{\textbf{One-magnon Raman responses.} Scattering intensity in the parallel channel ($\mathbf{e}_{\mathrm{in}}=\mathbf{e}_{\mathrm{out}}=[110]$) computed from the minimal model (\ref{smodel}). The insets show the polarization angular dependence of two modes in the parallel (red) and cross (black) channels, respectively.}
	\end{figure*}

To assign the Raman active phonon modes to their respective symmetries, and to highlight their anomalous behavior, we plot color contour maps of the angular dependent Raman response in parallel and crossed configurations for two different temperatures in Supplementary Figure 3. Excitations of all three symmetries $A_{\mathrm{1g}}$, $E_{\mathrm{g}}$, and $T_{\mathrm{2g}}$ are clearly distinguished at 300 K, following their expected behavior. One $T_{\mathrm{g}}$ phonon remains obscured either due to a weak scattering intensity, or due to a (partial) overlap with another phonon mode. At 5 K (i.e., below $T_{\mathrm{N}}$) the $E_{\mathrm{g}}$ phonon deviates from its four-fold symmetry (most clearly observed for \textbf{e}$_{\mathrm{in}}$ // \textbf{e}$_{\mathrm{out}}$) and instead follows a distorted two-fold pattern. Strikingly, the newly emerged 3-meV excitation mimics this behavior. We note that the maximal subgroup of $Fd\bar{3}m$ (No. 227) which can support such behavior is $Fd\bar{3}$ (No. 203). Even though the lowering of the symmetry increases the number of Raman-active modes to $A_{\mathrm{g}}$ + 2$E_{\mathrm{g}}$ + 6$T_{\mathrm{g}}$, the Raman tensors for $Fd\bar{3}$ resemble the exact same form as those for $Fd\bar{3}m$ except for the $E_{\mathrm{g}}$ channels.

Hence at $T=5$ K, all but $E_{\mathrm{g}}$ phonon modes retain their selection rules. At $T = 5$ K we clearly identify $A_{\mathrm{g}}$ (at 61.4 meV) and split $E_{\mathrm{g}}$ (at $\omega_1=37.8$ meV and $\omega_2=41.5$ meV) modes, as well as three out of the six $T_{\mathrm{g}}$ phonons (at 34.9 meV, 51.6 meV, and a faint one at 81.9 meV. Two additional candidates for $T_{\mathrm{g}}$ phonons might be located around 150 meV, not shown here).

\textbf{Supplementary Note 3 $|$ Magnetic model for one-magnon excitations}

Since peaks at $25$ meV and $32$ meV appear below the ordering temperature of the magnetic moments on Ru ions, we assume that these are magnetic excitations associated with this ordering. Here we adopt an oversimplified view of small spin-orbit coupling, in which one can start with the model of interaction $S=1$ spins despite the experimentally observed reduced value of 1.18 $\mu_{\mathrm{B}}$ for Ru magnetic moments. Under these assumptions, we use a generic super-exchange Hamiltonian of $S=1$ system on the pyrochlore lattice proposed in Refs.~\cite{li-18, gao-20}:

	\begin{align}\label{smodel}
		\mathcal{H}_{\mathrm{Ru}}=\sum_{\langle ij\rangle}\left[J\,\mathbf{S}_i\cdot\mathbf{S}_j+\mathbf{D}_{ij}\cdot\left(\mathbf{S}_i\times\mathbf{S}_j\right)\right]+\sum_{i}D_z\,(\mathbf{S}_i\cdot\hat{n}_{[111]})^2,
	\end{align}

where the vector $\mathbf{D}_{ij}$ denotes bond-dependent Dzyaloshinskii-Moriya (DM) interaction, and $D_z$ denotes the strength of the single-ion anisotropy along the local $[111]$ axis direction on each site. The phase diagram of this model  for a reasonable set of the model parameters has five $\bf{k}=0$ magnetic ground states:
all-in-all-out state, splayed ferromagnet, coplanar XY antiferromagnet (1) and (2), and non-coplanar XY antiferromagnet. We analyzed all of these configurations and concluded that the best description of the experimental data is achieved for the all-in-all-out ordered state of Ru magnetic moments. The linear spin wave analysis in the all-in-all-out state gives two modes at $\bf{k}=0$:
	\begin{eqnarray}\label{eqs}
		&&E_1=6\sqrt{2}D-D_z,\\
		&&E_2=\sqrt{\frac{56D^2+D_z(3D_z-16J)+4\sqrt{2}D(8J-5D_z)}{3}},\nonumber
	\end{eqnarray}
	where $D=|\mathbf{D}_{ij}|$, $E_1$ is non-degenerate, and $E_2$ is three-fold degenerate. These modes can give rise to the one-magnon Raman response.	Since we observe that the $32$ meV mode shows stronger intensity compared to the $25$ meV mode (Supplementary Figs. 3c and 3d), we let $E_2=32$ meV and $E_1=25$ meV. Solving Eq.(\ref{eqs}) gives us analytical expressions for $J$ and $D_z$ in terms of $D$. By setting DM interaction to a reasonable value $D=2.0$ meV, we get other parameters to be equal to $J=10.05$ meV and $D_z=-8.03$ meV.

Then we use the Loudon-Fleury (LF) form \cite{Loudon68} for the Raman operator 
		\begin{align}
		\mathcal{R}_{ij}=(\mathbf{e}_\mathrm{in}\cdot\mathbf{r}_{ij})(\mathbf{e}_\mathrm{out}\cdot\mathbf{r}_{ij})\left[J\,\mathbf{S}_i\cdot\mathbf{S}_j+\mathbf{D}_{ij}\cdot\left(\mathbf{S}_i\times\mathbf{S}_j\right)\right],
	\end{align}
		where $\mathbf{r}_{ij}$ is the vector indicating the bond $\langle i,j \rangle$, and $\mathbf{e}_{\mathrm{in}}$ ($\mathbf{e}_{\mathrm{out}}$) denotes the polarization of the incident (outgoing) light, to compute the one-magnon Raman response. (Note that the single ion anisotropy term $D_z$ does not give a contribution to the LF one-magnon Raman operator.) The resulting Raman response in the parallel polarization channel ($\mathbf{e}_{\mathrm{in}}=\mathbf{e}_{\mathrm{out}}=[110]$) is shown in Supplementary Figure 4. The inset shows the polarization angular dependence of the  two modes in the parallel (red) and cross (black) channel. The obtained angular polarization dependance of the Raman response at $32$ meV is in a good agreement with experimental data shown in Supplementary Figures 3c and 3d. However, the angular polarization dependence of the $25$ meV mode differs from the one observed experimentally. There are several possible factors that complicate the comparison between theory and experiment: the picture used in our theoretical analysis may have been oversimplified, such as ignoring other anisotropic interactions and non-Loudon-Fleury processes in calculating Raman response~\cite{yang-21}. Experimentally, there is also an overall inhomogeneous, noisy scattering background, which makes it difficult to determine the polarization dependence of the weak 25 meV mode. However, this will not qualitatively change our picture that both $25$ meV and $32$ meV modes come from the one-magnon excitations.
	
\textbf{Supplementary Note 4 $|$ Symmetry elimination of one-magnon origin of the $A$-mode}

In this section we show that the $A$-mode {\it can not} be understood as a one-magnon Raman response in any of the magnetically ordered phases of the minimal $S=1$ model Eq.(\ref{smodel})~\cite{li-18, gao-20}. 
 To compute the Raman response, we again employ a LF approach \cite{Loudon68}. The symmetry consideration allows us to decompose the Raman operator into different symmetry channels according to the irreducible representation of the point group symmetry of the ground state: $\mathcal{R}_{\mathrm{LF}}=\sum_{\langle i,j\rangle}\mathcal{R}_{ij}=\sum_{\alpha,\beta=x,y,z}\xi^{\alpha\beta}\left(\sum_{\langle i,j\rangle}\mathcal{R}^{\alpha\beta}_{ij}\right)=\sum_{\Gamma}\xi^{\Gamma}\mathcal{R}^{\Gamma}$, where $\xi^{\alpha\beta}\equiv(\textbf{e}_{\mathrm{in}}^\alpha \textbf{e}_{\mathrm{out}}^\beta+\textbf{e}_{\mathrm{in}}^\beta \textbf{e}_{\mathrm{out}}^\alpha)/2$, $\mathcal{R}^{\alpha\beta}\equiv\sum_{\langle ij\rangle}\mathcal{R}^{\alpha\beta}_{ij}\equiv \sum_{\langle ij\rangle}\textbf{r}_{ij}^{\alpha}\textbf{r}_{ij}^{\beta}\mathcal{H}_{ij}$, and $\Gamma$ labels the irreducible representation. Since the Raman response is computed by $I(\omega)=\int dt\, \mathrm{e}^{\mathrm{i} \omega t}\langle \mathcal{R}_{\mathrm{LF}}(t)\mathcal{R}_{\mathrm{LF}}(0)\rangle$, and the grand orthogonality theorem gives $\langle \mathcal{R}^{\Gamma_\mu}(t)\mathcal{R}^{\Gamma'_\nu}(0)\rangle\propto\delta_{\Gamma\Gamma'}\delta_{\mu\nu}$, the product of Raman operators from different irreducible representations gives no Raman response.

Now we can analyze the symmetry of the Raman response state by state. The all-in-all-out state has $T_{\mathrm{d}}$ point group, thus the Raman operator can be decomposed into $A_1$, $E$ and  $T_{2}$ irreducible representations as
		\begin{align}
			I(\omega)\propto&(\xi^{A_1})^2\langle \mathcal{R}^{A_1}(t)\mathcal{R}^{A_1}(0)\rangle+(\xi^{E^{(1)}})^2\langle \mathcal{R}^{E^{(1)}}(t)\mathcal{R}^{E^{(1)}}(0)\rangle\nonumber\\+&(\xi^{E^{(2)}})^2\langle \mathcal{R}^{E^{(2)}}(t)\mathcal{R}^{E^{(2)}}(0)\rangle+
			(\xi^{T_{2}^{(1)}})^2\langle \mathcal{R}^{T_{2}^{(1)}}(t)\mathcal{R}^{T_{2}^{(1)}}(0)\rangle.
			\label{eqn:Td}
		\end{align}
		For comparison with the polarization dependence shown in the insets of Figs.~1a and 1b of the main text, we set $\textbf{e}_{\mathrm{in}}^{||}=(\cos\theta,\sin\theta,0)$, $\textbf{e}_{\mathrm{out}}^{||}=(\cos\theta,\sin\theta,0)$ for the parallel channel, and $\textbf{e}_{\mathrm{in}}^{\perp}=(\cos\theta,\sin\theta,0)$, $\textbf{e}_{\mathrm{out}}^{\perp}=(-\sin\theta,\cos\theta,0)$ for the crossed channel. Without loss of generality, we denote $\langle\mathcal{R}^{A_1}(t)\mathcal{R}^{A_1}(0)\rangle=r_1$, $\langle\mathcal{R}^{E^{(1)}}(t)\mathcal{R}^{E^{(1)}}(0)\rangle=\langle\mathcal{R}^{E^{(2)}}(t)\mathcal{R}^{E^{(2)}}(0)\rangle=r_2$, and $\langle \mathcal{R}^{T_{2}^{(1)}}(t)\mathcal{R}^{T_{2}^{(1)}}(0)\rangle=r_{3}$, where $r_1,r_2$ and $r_3$ are real numbers. Clearly, only one of $r_1$, $r_2$ and $r_3$ is non-zero when considering the Raman response for a given one-magnon eigenmode, since it can only belong to one irreducible representation. Then the Raman responses in the parallel and in the crossed channel are given by
		\begin{align}
			I^{||}(\omega)\propto&\,r_1+\frac{r_3}{8}+\frac{11}{72}r_2+\frac{1}{8}(r_2-r_3)\cos 4\theta\label{eq:T2g_para_sim},\\
            I^{\perp}(\omega)\propto&\,\frac{1}{8}(r_3+r_2)+\frac{1}{8}(r_3-r_2)\cos4\theta\label{eq:T2g_perp_sim}.
		\end{align}
  As expected, if the mode belongs to the $A_1$ irreducible representation ($r_1\neq 0$), there is no angular dependence. If the mode is either in $E$ or in $T_{2}$ irreducible representation, then because of the $\cos4\theta$ angular dependence, both $I^{||}(\omega)$ and $I^{\perp}(\omega)$ always preserve the four-fold symmetry pattern. This excludes the possibility that the $A$-mode is due to the one-magnon Raman scattering in the all-in all-out state.
		
		For the remaining four spin configurations, the Raman operator can be decomposed according  to the irreducible representation of the $D_{\mathrm{2d}}$ point group.
		 Denoting  $\langle\mathcal{R}^{A_1}(t)\mathcal{R}^{A_1}(0)\rangle=r_1'$, $\langle\mathcal{R}^{B_1}(t)\mathcal{R}^{B_1}(0)\rangle=r_2'$, and $\langle\mathcal{R}^{B_2}(t)\mathcal{R}^{B_2}(0)\rangle=r_3'$,  we obtain the Raman response in the parallel and in the crossed channel as
		\begin{align}
			I^{||}(\omega)\propto&\,\frac{1}{8}(8r_1'+4r_2'+r_3')+\frac{1}{8}(4r_2'-r_3')\cos4\theta,\label{eqn:para_D2d}\\
		    I^{\perp}(\omega)\propto&\,\frac{1}{8}(4r_2'+r_3')+\frac{1}{8}(-4r_2'+r_3')\cos4\theta.\label{eqn:cros_D2d}
		\end{align} 
		The four-fold symmetry pattern in the polar plot is also preserved here because of $\cos 4\theta$, thus excluding the possibility that the $A$-mode is due to the one-magnon Raman scattering in all these states.

\textbf{Supplementary Note 5 $|$ Raman tensor for the Higgs-type amplitude mode $A$}

The polarization-dependent Raman intensity of the amplitude mode can be fully described with the following Raman tensor, as demonstrated in Supplementary Figure 5:

\begin{figure}
\label{figure7}
\centering
\includegraphics[width=6cm]{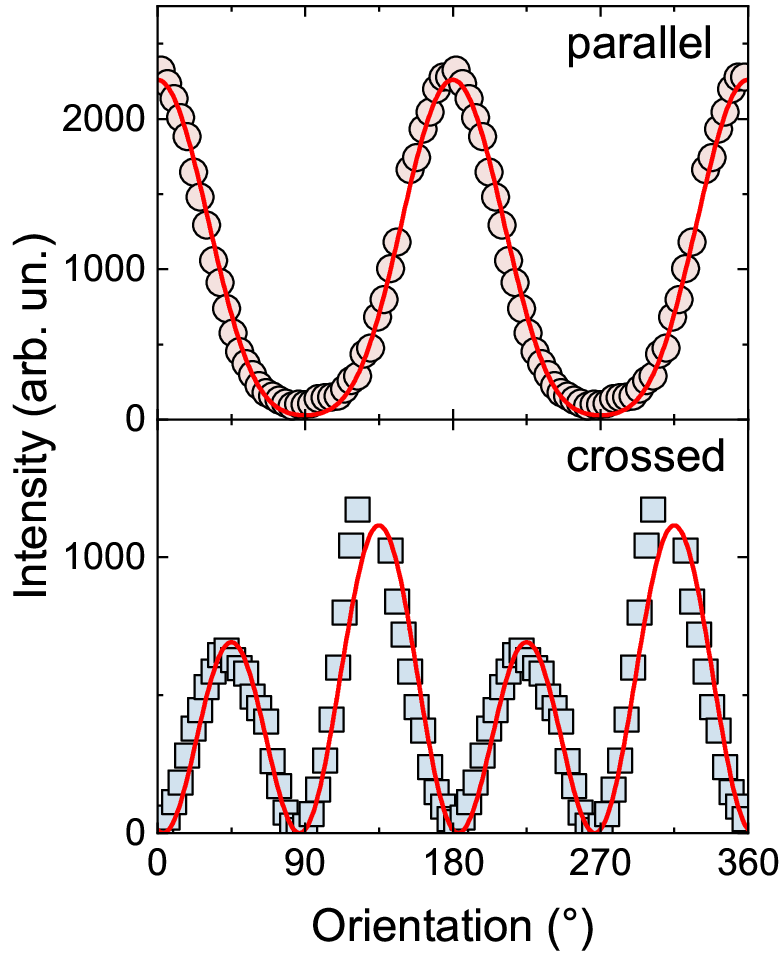}
\caption{\textbf{Polarization-dependent Raman scattering intensity of the amplitude mode (symbols) fitted with Raman tensor $R_{\mathrm{Amp}}$ (solid lines).}}
\end{figure}

\begin{center}
	\mbox{$R_{\mathrm{Amp}}$=$\begin{pmatrix} a & d & ...\\ -d & b & ...\\ ... & ... & ...\\ \end{pmatrix}$}.
	\end{center}

As our measurements have been carried out exclusively within the crystallographic $ab$-plane, we can only comment on Raman tensor elements relevant to this plane. These tensor elements (as well as the observed polarization-dependence) are also consistent with the previously reported CDW mode symmetry of GdTe$_3$~\cite{wang-22}. In both cases, Nd$_2$Ru$_2$O$_7$ and GdTe$_3$, the amplitude modes show a two-fold symmetry in crossed polarization, therefore antisymmetric off-diagonal tensor elements are required. To achieve a full description of the Raman tensor, including $c$-axis tensor elements, future experiments refining the Ru-spin structures below $T_{\mathrm{N}}$ as well as below $T^*$ will be crucial.

\textbf{Supplementary Note 6 $|$ Fitting of the low-energy spectral range}

Supplementary Figure 6 compares different fitting approaches to the low-energy spectral range, dominated by the quasi-elastic scattering (QES) and the amplitude excitation ($A$-mode): Above $T_{\mathrm{N}}$, the spectral weight in the as-measured data towards 0 meV remains insignificant. Below $T_{\mathrm{N}}$, spectral weight builds up rapidly with decreasing temperature. This spectral weight can be fitted well with a single Lorentzian line centered at 0 meV (here the imposed fitting constraint is that the line's energy is fixed at 0, while linewidth and intensity are free fitting parameters). The ``single quasi-elastic-line'' approach yields satisfactory results down to about 110 K. At 85 K and below a second line centered at finite energies is required to yield a reasonable description of the data. For this second line no fitting constraints were set. The cross-over from dominating quasi-elastic to finite-energy mode, far below $T_{\mathrm{N}}$, is also directly observed in the color-contour plot of the measured Raman data in Fig. 1(c). Fits to the data at 95 K (i.e., close to the cross-over temperature $T^*$) are ambiguous and both approaches -- with and without a second line -- can aptly describe the as-measured data. Error bars in Figs. 1e-g reflect those ambiguities.

\begin{figure*}
\label{figure8}
\centering
\includegraphics[width=16cm]{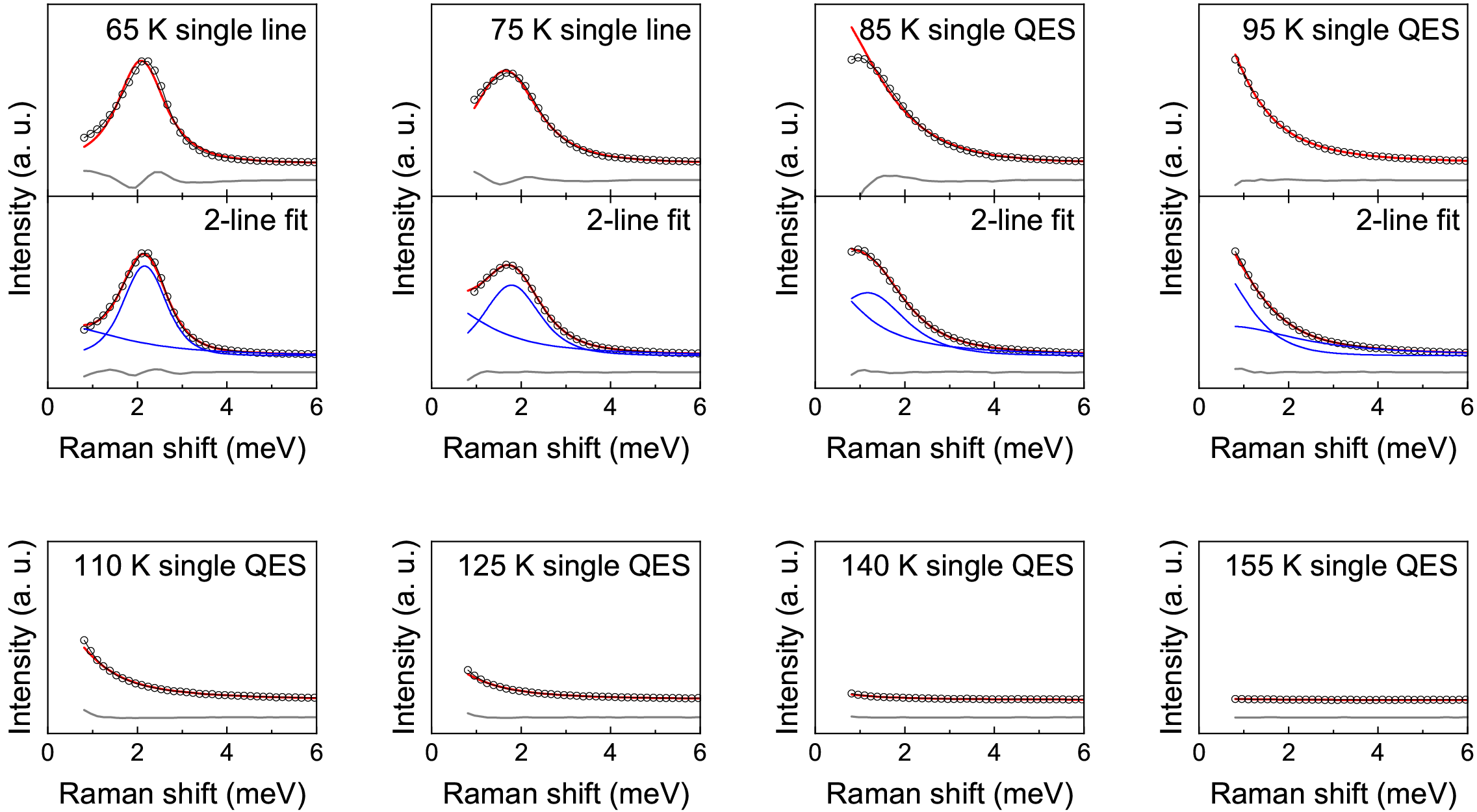}
\caption{\textbf{Fits to the as-measured low-energy Raman scattering intensity at eight different temperatures.} The solid red lines denote the (sums of) fits and blue lines indicate fits to individual components, open symbols are as-measured data, and solid gray lines represent the difference between fit and data.}
\end{figure*}

\textbf{Supplementary Note 7 $|$ Temperature dependence of one-magnon modes}

\begin{figure*}
\label{figure9}
\centering
\includegraphics[width=6cm]{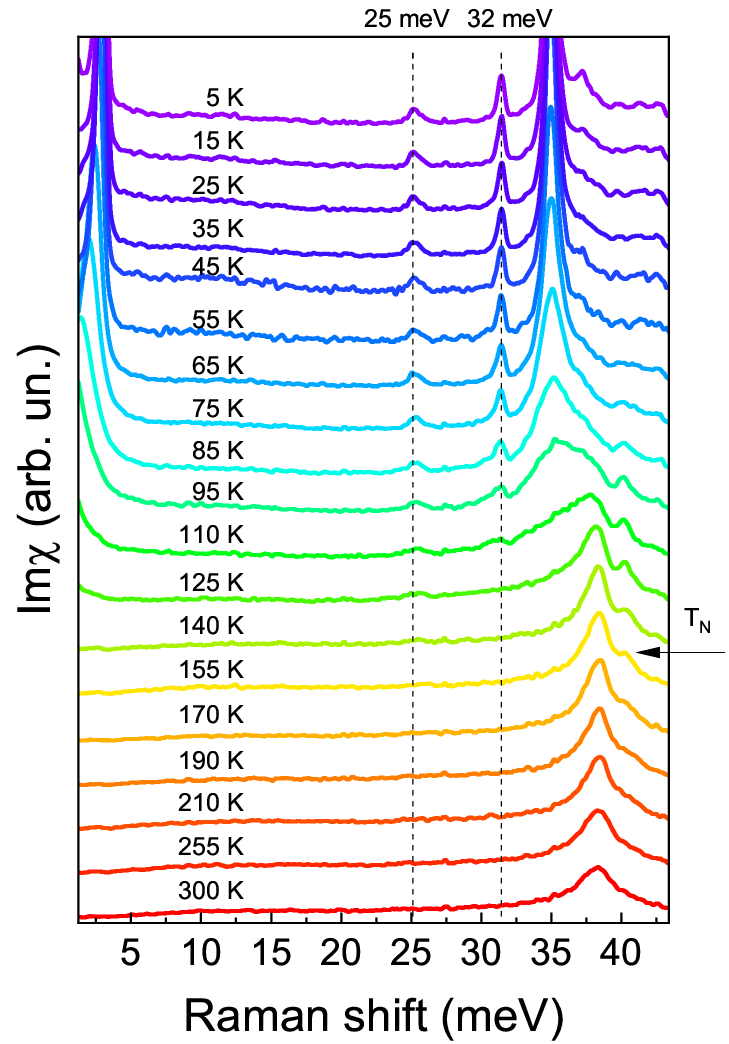}
\caption{\textbf{Temperature dependent Raman spectra.} Data recorded in parallel light polarization, containing the $A_{\mathrm{1g}}$ + $T_{\mathrm{2g}}$ symmetry channel.}
\end{figure*}

The full temperature dependence of the low-energy spectral range is displayed in Supplementary Figure 7, focusing on the evolution of one-magnon excitations at 25 meV and 32 meV (marked by dashed black lines). In contrast to the $A$-mode at 3 meV, these two excitations show no discernible energy- or linewidth-dependence on temperature, and gradually decrease in intensity upon approaching $T_{\mathrm{N}}$. Note that the absence of a softening while approaching $T_{\mathrm{N}}$ can be understood by considering their sizable spin gaps with respect to the transition temperature. In this case, quantum- and thermal fluctuations which could cause a redshift are partially suppressed.

\textbf{Supplementary Note 8 $|$ Phonon anharmonicity}

The temperature dependence of the phonon frequency $\omega(T)$ and linewidth $\Gamma(T)$ shown in Figs. 2c-d (main text) has been fitted by conventional cubic second-order anharmonic softening and broadening, respectively, as indicated by the solid red lines. The corresponding fitting functions are given by $\omega (T) = \omega_0 - C \left( 1+\frac{2}{\mathrm{exp}(\hbar \omega_0 / 2k_{\mathrm{B}} T) -1} \right)$, and $\Gamma (T) = \Gamma_0 \left( \frac{1+D}{\mathrm{exp}(\hbar \omega_0 / k_{\mathrm{B}} T) -1} \right)$, where $C$ and $D$ are fitting parameters~\cite{balkanski-83}.

\end{document}